\documentclass{article}

\usepackage{arxiv}

\usepackage[utf8]{inputenc} % allow utf-8 input
\usepackage[T1]{fontenc}    % use 8-bit T1 fonts
\usepackage{hyperref}       % hyperlinks
\usepackage{url}            % simple URL typesetting
\usepackage{booktabs}       % professional-quality tables
\usepackage{amsfonts}       % blackboard math symbols
\usepackage{nicefrac}       % compact symbols for 1/2, etc.
\usepackage{microtype}      % microtypography
\usepackage{graphicx}
\usepackage{doi}

\usepackage{newtxtext,newtxmath}
\usepackage{textcomp}
\usepackage{latexsym}

\usepackage{multirow}

\title{Differences between preprints and journal articles : \\Trial using bioRxiv data}

\author{KOSHIBA Hitoshi \\
	 NISTEP\thanks{National Institute of Science and Technology Policy}\\
	Chiyoda, Tokyo 100-0013 JAPAN \\
	\texttt{hkoshiba@ieee.org}
	\And
	HAYASHI Kazuhiro \\
	NISTEP\\
	Chiyoda, Tokyo 100-0013 JAPAN \\
	\texttt{khayashi@nistep.go.jp}
}

\renewcommand{\shorttitle}{Differences between preprints and journal articles}

\hypersetup{
pdftitle={Differences between preprints and journal articles},
pdfsubject={cs.DL},
pdfauthor={KOSHIBA etal},
pdfkeywords={Science Metrics, Preprint, Publication, bioRxiv, Scientific journal},
}

\begin{document}
\maketitle

\begin{abstract}

In this paper, we attempted to obtain knowledge about how research is conducted, especially how journal articles are produced, by comparing preprints with journal articles that are finally published.

First, due to the recent trend of open journals, we were able to secure a certain amount of full-text XML of preprints and journal articles, and verified the technical feasibility of comparing preprints and journal articles.
On the other hand, within the scope of this trial, in which we tried to clarify the difference between them based on external criteria such as the number of references and the number of words, and simple document similarity, we could not find a clear difference between preprints and journal articles, or between preprints that became journal articles and those that did not. Even with the machine learning method, the classification accuracy was not high at about 47\%.

The result that there is no significant difference between preprints and journal articles is a finding that has been shown in previous studies and has been replicated in larger and relatively recent situations.
In addition to these, the new findings of this paper are that the differences in many external criteria, such as the number of authors, are small, and the differences with preprints that are not journal articles are not large.
\end{abstract}

% keywords can be removed
\keywords{Science Metrics \and Preprint \and Publication \and bioRxiv \and Scientific journal}

% ==================================================================
\section{Introduction}

How is research carried out, for example, how is a journal article produced? Researchers have the answers to the above questions because they actually conduct research and generally summarize their findings in the form of papers. On the other hand, they are individual episodes, and it is not always clear whether they are common or not, or to what extent there are variations.

In recent years, digital transformation (DX) has been progressing, and various activities have come to be conducted mainly on the Internet. As a result, it has become possible to observe and analyze the processes of various social phenomena, which has been difficult in the past, and research fields such as computational social science are gaining momentum.

In the early days of the Internet, it was actively used for exchanging research information including papers, but the flow of DX described above has made it possible to visualize and analyze the research process. For example, as of 2021, in the fields of physics and information science, the research style of publishing preprints, which are drafts of papers before submission to journals, through preprint servers has become widespread. In addition, with the worldwide spread of COVID-19, the research style through preprint servers is also becoming popular in the fields of medicine and biology. This has made it possible to examine the differences between the stage before submission as a journal article and the stage when the article is finally turned into a journal article, as well as the differences between preprints that have been turned into journal articles and those that have not yet been turned into journal articles. It is now possible to examine the differences between preprints that have become journal articles and those that have not yet become journal articles.

In this paper, we report the results of a survey on the differences between the journal articles and their preprints, and between preprints that have become journals and those that have not, using bioRxiv, a preprint server for biological systems, with the ultimate goal of understanding how research is conducted.

% ==================================================================
\section{Related Works}
Similar previous studies include Kleinll,Carneiro,Akbaritabar \cite{Klein2019,Carneiro2020,Akbaritabar2021}.

Klein\cite{Klein2019} compares journal articles with their preprints, which is very similar to the content of this paper.
Klein compared more than 12,000 preprints and their journal versions on arXiv and bioRxiv, and reported that there was little difference between them.
In the case of bioRxiv, the analysis covered the period from the site's launch in November 2013 to November 2016, with a total of 7,000 preprints and a relatively small number of full-text articles (220).
For the content, we adopted the Bag-of-Words Cos similarity, where each word is an independent dimension.
As described below, this paper is limited to bioRxiv and the period is 2019. The number of preprints is more than 20,000 in total, and about 7,000 full-text articles are used, which means that the situation in bioRxiv is relatively recent and more extensively investigated.
As for the content, the similarity is judged based on distributed representation.

Carneiro\cite{Carneiro2020} also analyzed the change in quality of journal articles that were published in bioRxiv and later in PubMed, and reported that the quality of peer-reviewed journal articles was slightly higher than that of preprints, but the difference was not significant.
The Carneiro's paper, bioRxiv covers the papers submitted in 2016, and the change in quality is mainly based on questionnaires.
Since it is a questionnaire, the number of pairs of preprints and journal articles used for the analysis is 56 cases.
The difference between this paper and the Carneiro's is that, Carneiro's is the survey period is relatively old (2016), and a small number of data are surveyed on a questionnaire basis.

Akbaritabar\cite{Akbaritabar2021} investigates the difference in references between preprints and journals.
In addition to quantitatively examining more than 6000 pairs of preprints and journal articles, he also evaluates about 100 of them by human readers to classify the types of context added and to identify differences by field.
The difference is that the paper focuses on references and attempts a precise analysis, while this paper provides a comprehensive overview of various indicators.

% ==================================================================
\section{Data Description}

In this section, we will discuss the analysis data.

% ------------------------
\subsection{Preprint Server; bioRxiv}
This time, we set bioRxiv, which is used in the field of biological science, as the preprint server. The reasons are described below.

As a preprint server, arXiv, which is often used in the fields of physics and information science, is well known, has the longest history, and has a large number of submissions and downloads\footnote{For example, the number of submissions and downloads can be checked below. \url{https://arxiv.org/help/stats} }. 
From this point of view, arXiv is considered to be appropriate. However, in the field of information science, journal papers are not necessarily important, and the proceedings of top conferences have the same value as journal papers, which complicates the discussion when we consider comparison with journal papers. In addition, although the sources of the manuscripts are available in TeX format, analysis is not always easy because of the high degree of freedom in customization. In addition, it is not clear whether the sources of journal papers and proceedings of top conferences are available or not, or whether they can be obtained in HTML or XML format for easy analysis.

On the other hand, in fields other than information science, it has not been observed that the proceedings of top conferences are as valuable as journal articles. In addition, bioRxiv provides full-text XML in JATS (Journal Article Tag Suite) format. The journal articles that were accepted via bioRxiv are also in many open access articles, and many of them can be obtained in full-text XML in JATS format as described below. We selected bioRxiv because the final results are expected to be linked to journal articles, and the data of preprints and journal articles can be obtained in the same format. The disadvantage of bioRxiv is that the number of analysis targets is less than 10,000 due to the difference in diffusion rate.

% ------------------------
\subsection{Target of analysis (period), number of data}

This time, we chose preprints submitted in 2019 as the target (period) for analysis. The reasons are as follows.

First, the field of biological science is also closely related to COVID-19, and a certain number of COVID-19-related contributions have been observed since 2020. Although these numbers are not large in the context of bioRxiv as a whole, they may act as a disturbance and complicate the discussion. Second, using older data can eliminate these effects, but if the data is too old, the possibility that it is not applicable to the current situation increases due to changes in trends.

Here, COVID-19-related preprints started to increase around the end of December 2019\cite{koshiba2020}, so using the 2019 data is relatively recent and does not take into account the impact of COVID-19. In addition, as previously reported, those papers submitted to bioRxiv that are accepted for publication in journals are turned into journal articles within 6 to 8 months on average from the time of preprint registration\cite{Abdill2019,hayashi2021}. In this paper, the analysis was conducted in May 2021, which is almost 17 months from the end of December 2019, and we can expect that the journal culture of the preprints registered in 2019 is almost complete and not many of them will be converted into journal papers in the future.

The number of journal articles with full-text XML for this analysis is shown below.

\begin{description}
\item[Number of journal preprints] 28,805.
\item[Number of journal articles] 13,450.
\item[Number of full-text XML holdings] 7,985.
\end{description}

Although preprints can be revised as many times as needed, only the first edition is considered in this analysis. This is due to the fact that the first edition of all submitted preprints definitely exists and is easy to control. If the closest edition from the journal article is used, the discussion becomes more complicated, such as whether to look at only the latest edition even if it has not been published in the journal, how to handle the case where there is an update after publication in the journal, and whether to look at the difference between the first edition and the latest edition.

% ------------------------
\subsection{Data acquisition method}

The data were obtained using the following methods.

For bioRxiv, detailed information of each preprint was obtained through bioRxiv API
\footnote{\url{https://api.biorxiv.org/}} in the past (April 2021). Since the detailed information includes``jats xml path'', we simply retrieved the information appropriately from those URLs. There are several hundred URLs that return an error when accessed, and these are excluded from the analysis.

For journal articles, there are several steps to follow.

First, the linkage between a preprint and a journal article is as follows. The aforementioned detailed information of each preprint contains ``doi'', and the DOI details of the preprint are obtained through the CrossRef API \footnote{\url{https://api.crossref.org/}}. 
Then, there is an attribute ``is-preprint-of'', and the value is set when there is a journal article based on the preprint. In this case, we used the case where the ``id-type'' of ``is-preprint-of'' is ``doi'' and a specific ID (DOI) is set. Therefore, the journal article in this paper is preprint-of. Therefore, all journal articles in this paper have a DOI.

Next, we obtain the full-text XML of journal articles in two steps. The first step is to obtain the full-text XML of journal articles in two steps. For many open access journals, PubMed Central (PMC) provides full-text XML in JATS format centrally in the PMC Open Access Subset \footnote{\url{https://www.ncbi.nlm.nih.gov/pmc/tools/openftlist/}}. 
In this section, we collect data using PubMed Central OAI-PMH (Open Archives Initiative Protocol for Metadata Harvesting) \footnote{\url{https://www.ncbi.nlm.nih.gov/pmc/tools/oai/}}.

PMC OAI-PMH collects data based on PMCIDs, which are IDs uniquely assigned to each article by PMC, and converts DOIs to PMCIDs through PMC's ID Converter API \footnote{\url{https://www.ncbi.nlm.nih.gov/pmc/tools/id-converter-api/}}. 
Then, the data is retrieved from OAI-PMH. Not all DOIs can be converted to PMCIDs, and the existence of PMCIDs does not necessarily mean the existence of full-text XML, but since it is difficult to confirm the existence of full-text XML in advance, we obtained data from OAI-PMH whenever PMCIDs were obtained.

% ==================================================================
\section{Methods}

This paper describes the analysis method.

% ------------------------
\subsection{Object of comparison}

We set up three patterns for comparison.

First, we compare the preprint that became a journal (journal preprint) with the journal, second, we compare the journal preprint with another similar preprint, and finally, we compare a randomly combined pair of preprints and journals as a baseline.

``Similar preprint'' requires some explanation and will be explained later.

% ------------------------
\subsection{Selection of feature values}

The following features were selected to observe the differences. Number of authors, number of references, number of paragraphs, number of words, chapter titles, and similarity of contents.

For the number of authors and references, in addition to a simple comparison of the number, the degree to which the names and titles match was also examined. For chapter titles, in addition to the number of matches, we also look at the number of words that are present in only one of the titles.

% ------------------------
\subsection{Concept of similarity}

The similarity of the contents described in the features is as follows.

Basically, we compare two documents (preprints and journal articles) on a paragraph-by-paragraph basis. In this case, the following steps should be taken to detect any change in the position of the paragraphs.

\begin{enumerate}
    \item detect the similarity between all paragraphs of the two documents A and B. 
    \item pair the paragraphs in B with the highest similarity from each paragraph in A.
    \item replace A and B and perform the process in 2 to create a pair seen from B. 
    \item take the pairs created by 2. and 3. as a whole, and make only the ones with a certain level of similarity as the common pairs of the two documents.
    \item calculate the Jaccard coefficient, which is the similarity between two documents A and B.
\end{enumerate}

The threshold value used in 4. is determined by looking at the distribution of the total similarity between the paragraphs calculated in 3.

In this specific work, each paragraph is converted into a vector using the variance representation, and the distance between the vectors is calculated. The direction is different in that the closer the distance, the more similar the paragraphs are, and the larger the similarity, the more similar the paragraphs are.

% ------------------------
\subsection{distributed representation}

In the aforementioned similarity calculation, it is common to calculate cosine similarity based on words, but this time we used variance representation. This was done in order to calculate the similarity by absorbing to some extent the differences in words and phrases even when they are changed in the process of peer review.

FastText\cite{Bojanowski2016}
\footnote{\url{https://fasttext.cc/}} 
was used as the specific method for obtaining distributed representation, and 300-dimensional embedding was performed in skip-grams using the data of preprints, journal article titles, summaries, and full-text data to be analyzed. Prior to the work, the data were Lemmatized and stopwords were removed based on NLTK and WordNet.

As a result, we obtained 300-dimensional distributed representations of 230,020 words. The distributed representation of paragraphs, etc., was obtained by linearly adding and normalizing the distributed representations containing them.

% ------------------------
\subsection{Methods for obtaining distributed representations}

In the field of biology, bioBERT\cite{Jinhyuk2019} \footnote{\url{https://github.com/dmis-lab/biobert}} is publicly available and can be used to obtain distributed representation with higher accuracy. However, for reasons of processing speed, we decided to forgo the use of bioBERT and use our own FastText-based distributed representation.

Specifically, we extracted five cases from the full-text data and calculated the processing speed. bioBERT took 27.8 seconds for a total of five cases, with an average of 5.6 seconds, while the FastText-based processing took 0.1 seconds for a total of five cases \footnote{In FastText, since the coordinate values do not change depending on the context
Since FastText does not change the coordinate values depending on the context, it stores the pre-calculated variance representation of each word in the DB, and reads and calculates it each time.}.

Since the FastText-based processing involves processing costs such as building a distributed representation dictionary, lemmatizing, and removing stop words, we should not use this calculation result alone to compare the costs. 
However, if we simply compare the processing speed per 5 sentences mentioned above, the difference is 278 times. Assuming that each document has 20 paragraphs, processing 20,000 documents would require 400,000 calculations, which is 622 hours (about 26 days) for bioBERT. On the other hand, the FastText-based method is expected to take about 2 hours.

Since the reason for using distributed representation is to absorb fluctuations in words and phrases, and not to pursue exact accuracy, we adopted FastText-based processing for its speed. The difference in accuracy between bioBERT and FastText-based processing is a future challenge.

% ------------------------
\subsection{The concept of ``similar preprints''}

When comparing journal preprints with other general preprints, it is easy to compare journal preprints with other statistics directly, but it is difficult to compare journal preprints with journals. Therefore, we search for preprints that are similar to journal preprints, and use the results of comparison with these similar preprints.

The ``similar preprint'' should be the one with the smallest distance between the variance representation values of the summary (the one with the largest similarity). In this case, the similar preprints are limited to those with older timestamps than the journal preprints and those that have not been journaled. This eliminates the possibility of being influenced by journals or journal preprints as well as being unjournaled.

For those with the smallest distance of variance, we used NGT \cite{Iwasaki2018}
\footnote{\url{https://github.com/yahoojapan/NGT}}
to approximate the top 100 preprints in order to reduce the computational complexity. There were about 1,000 cases where none of the top 100 similar preprints satisfied the aforementioned conditions (old in time and not journaled), resulting in 6,905 pairs of journal preprints and similar preprints. In addition, there are many cases where the same preprint is associated with multiple journal preprints, and the unique number of ``similar preprints'' is 3,874.

% ==================================================================
\section{Result}

% ------------------------
\subsection{Initial Setup}

\begin{figure}[htbp]
\centering
\includegraphics[width=70mm]{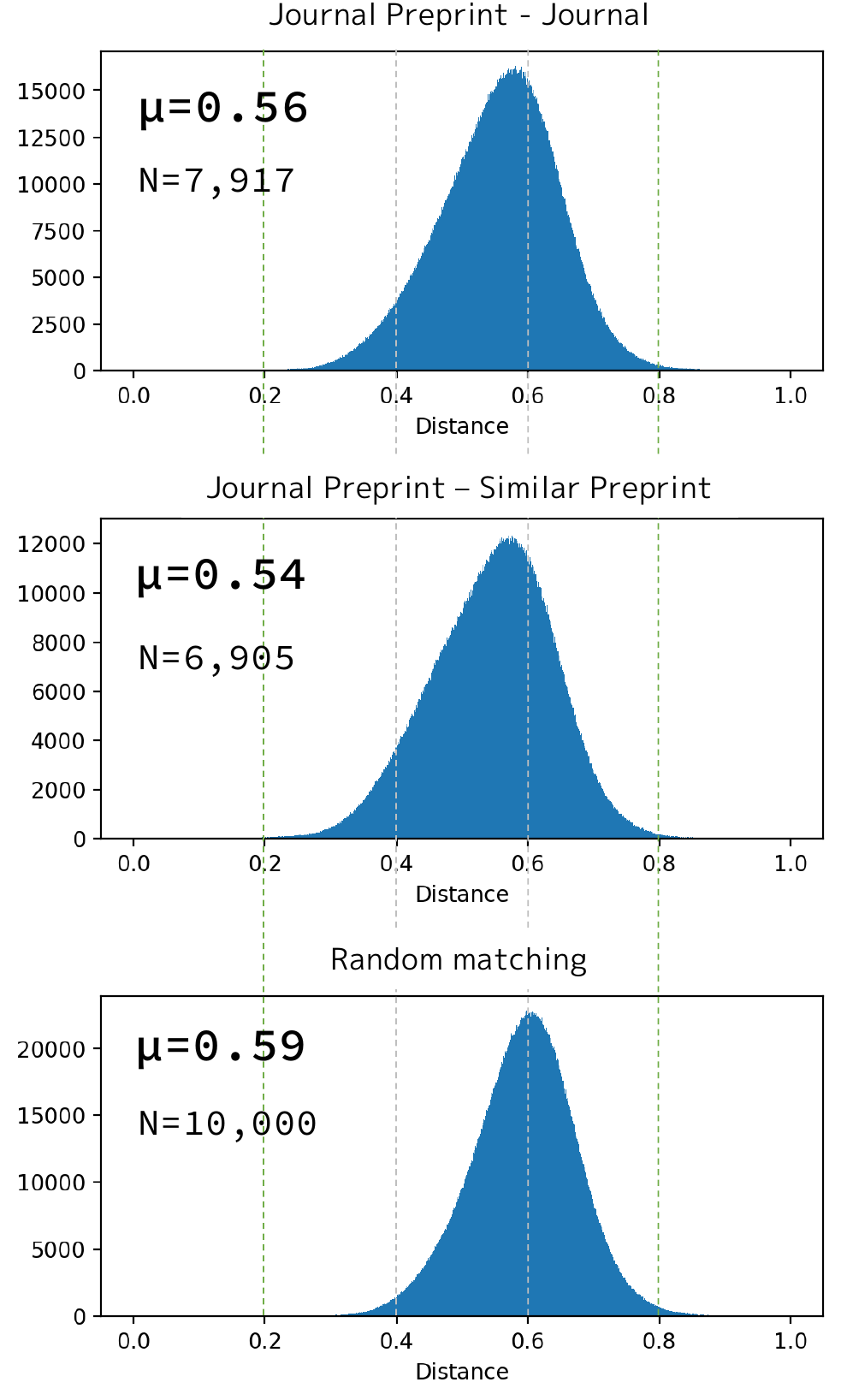}
\caption{Distribution of distances in all paragraph combinations}
\label{fig_th_dist}
\end{figure}

Figure \ref{fig_th_dist} shows the distribution of similarity (distance) for all paragraph combinations of journal preprints - journals, journal preprints - similar preprints, and random matching pairs, which are used to define the threshold for similarity calculation.

Figure \ref{fig_th_dist} shows that in the case of random matching, few distances below 0.3 are observed. On the other hand, in the case of journal preprints and the combination of journal and similar preprints, some of them are above 0.3. Therefore, we adopt a distance of 0.3 as the threshold for calculating the Jaccard coefficient between documents.

% ------------------------
\subsection{Similarity of content}

\begin{figure}[htbp]
\centering
\includegraphics[width=70mm]{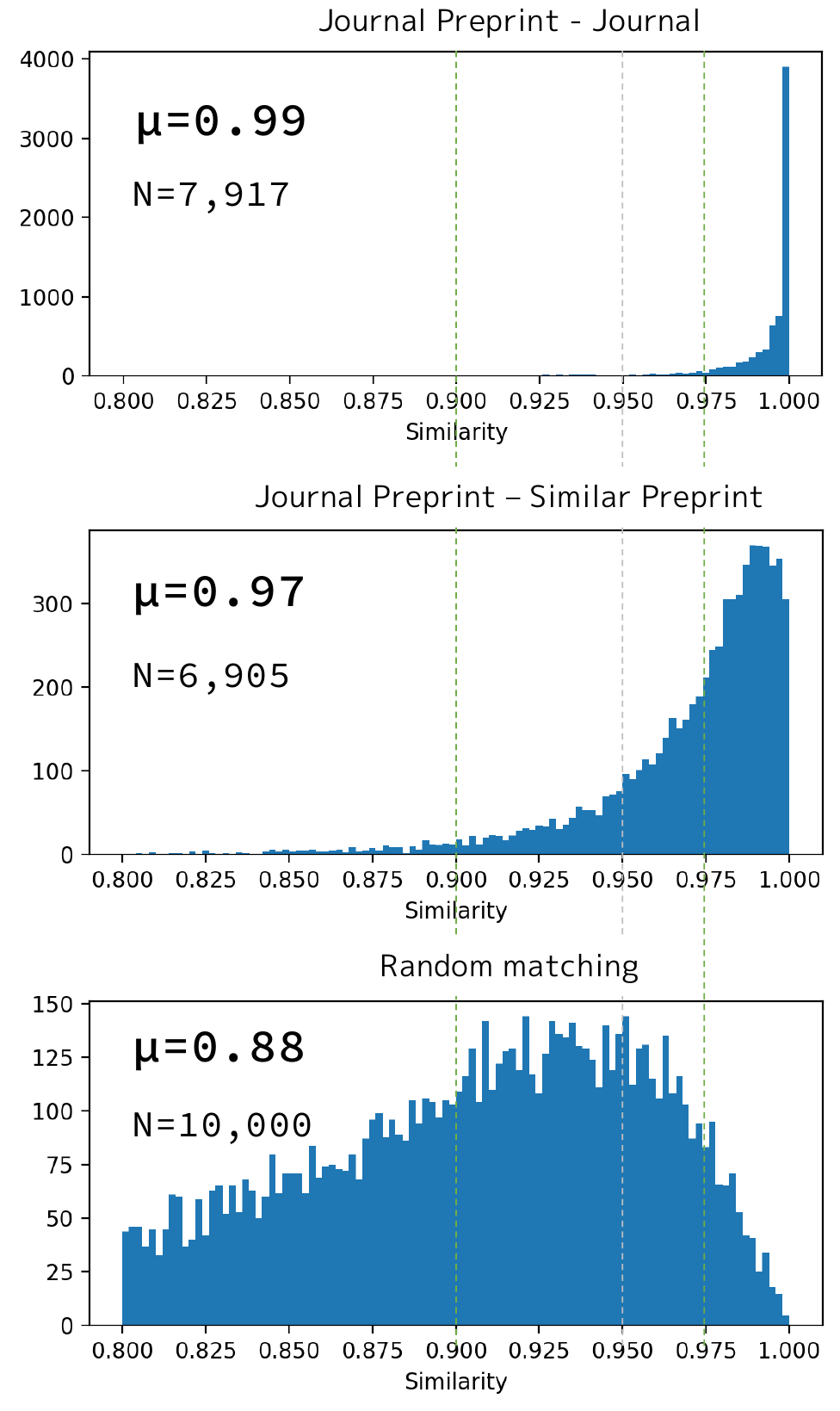}
\caption{Distribution of similarity between documents}
\label{fig_jaccard}
\end{figure}

Figure \ref{fig_jaccard} shows the distribution of Jaccard coefficients between documents calculated based on the above threshold.

Looking at Figure \ref{fig_jaccard}, first, the baseline random match has a distribution with few similarities above 0.95 and a peak around 0.93. Journal preprints and journal articles naturally have high similarity, peaking at 1.0. Some preprints that are similar to journal preprints show high similarity, such as 1.0, but the distribution is gentle, which is natural considering that the authors themselves are basically different.

% ------------------------
\subsection{Other external criteria}

External criteria other than the similarity of contents, such as the number of authors and the number of references, are shown below.

Figures \ref{fig_auth_ref} and \ref{fig_auth_ref_num} show the difference and the real distribution of the number of authors and the number of references. The distribution of the number of references shifts neatly from that of ``journal article'' to that of ``reference paper'', indicating that the number of references tends to increase when it becomes a journal article.

\begin{figure*}[htbp]
\centering
\includegraphics[width=140mm]{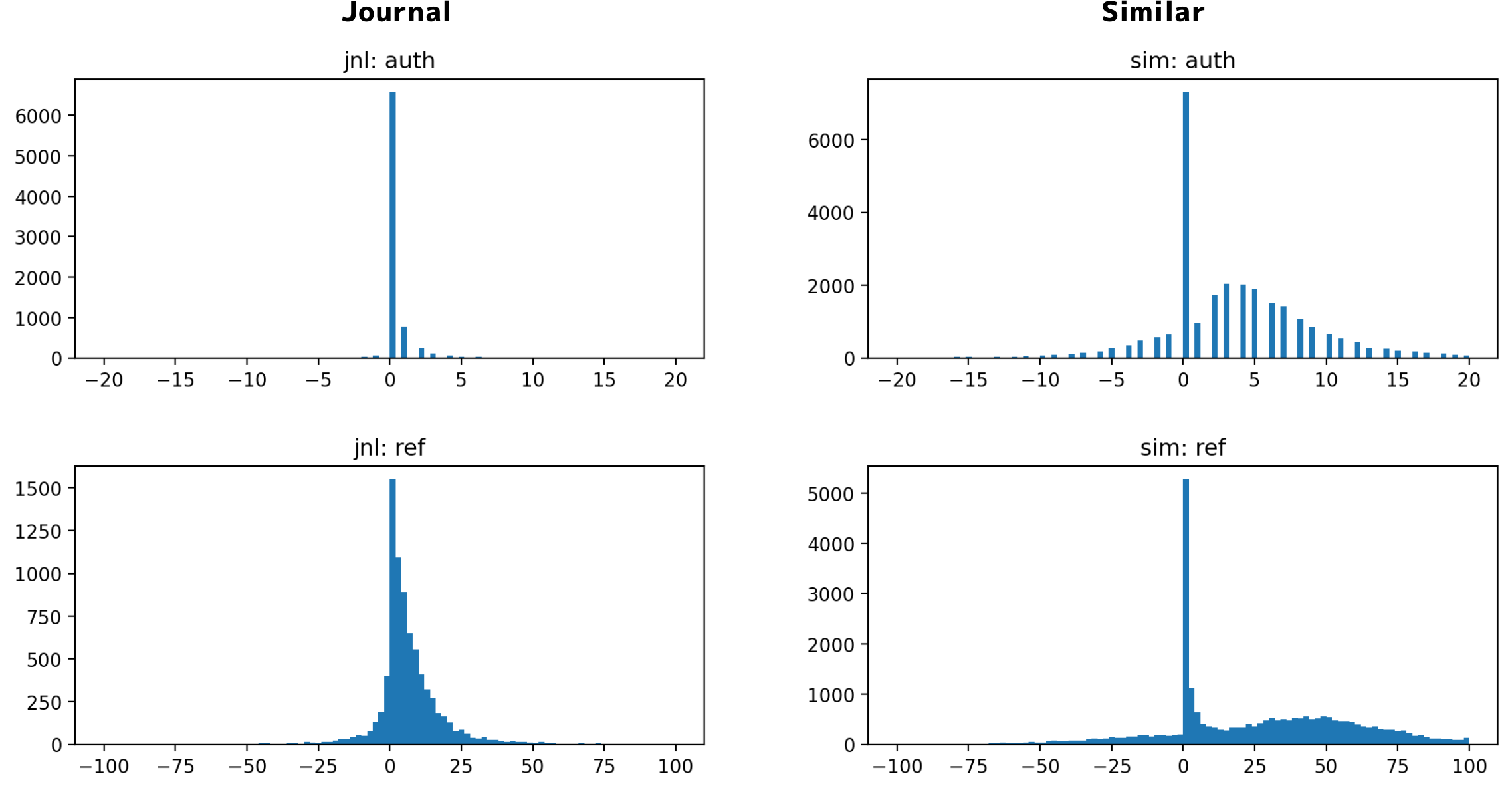}
\caption{Distribution of the difference in the number of authors and the number of references}
\label{fig_auth_ref}
\end{figure*}

\begin{figure*}[htbp]
\centering
\includegraphics[width=140mm]{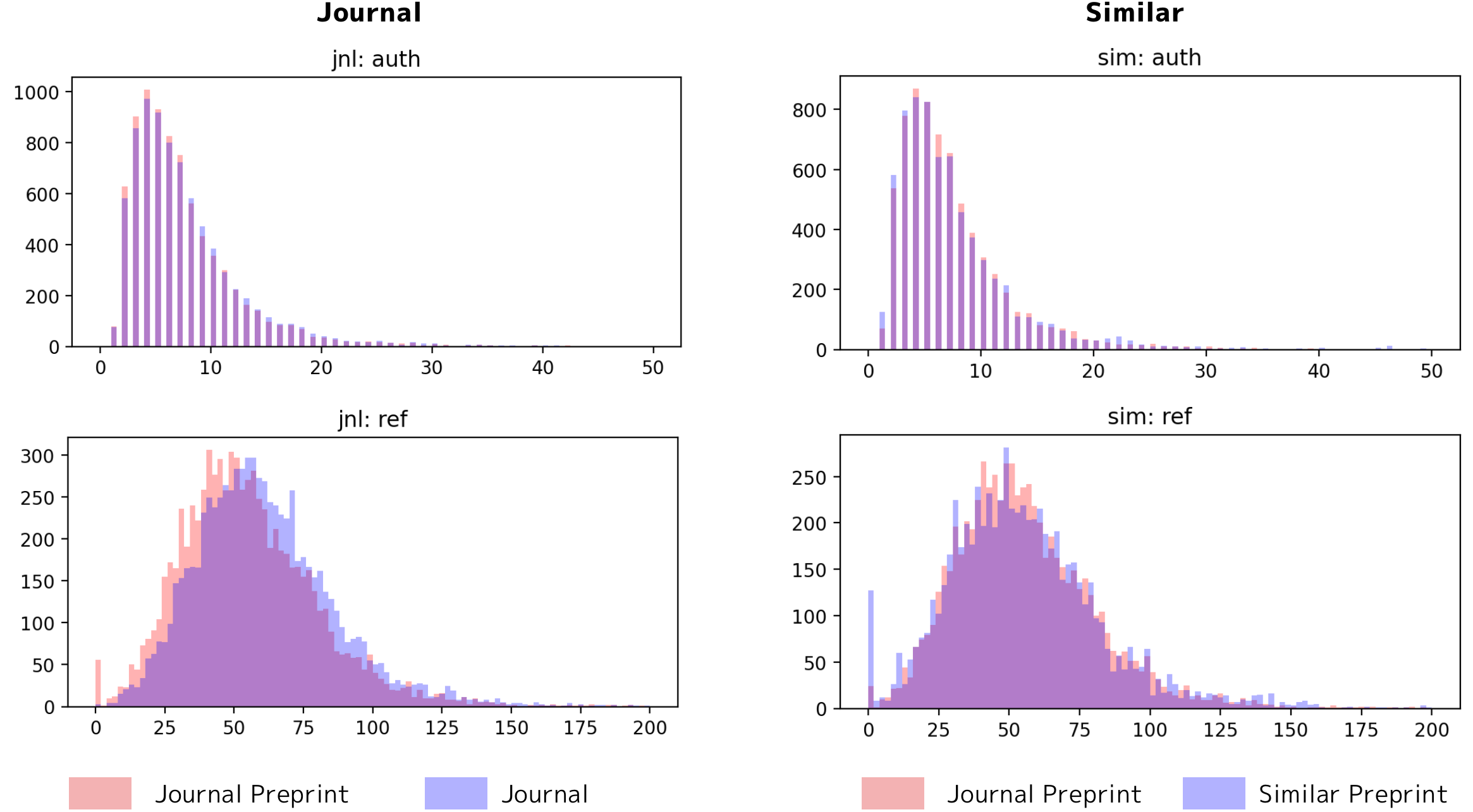}
\caption{Distribution of the number of authors and the number of references}
\label{fig_auth_ref_num}
\end{figure*}

\begin{figure*}[htbp]
\centering
\includegraphics[width=140mm]{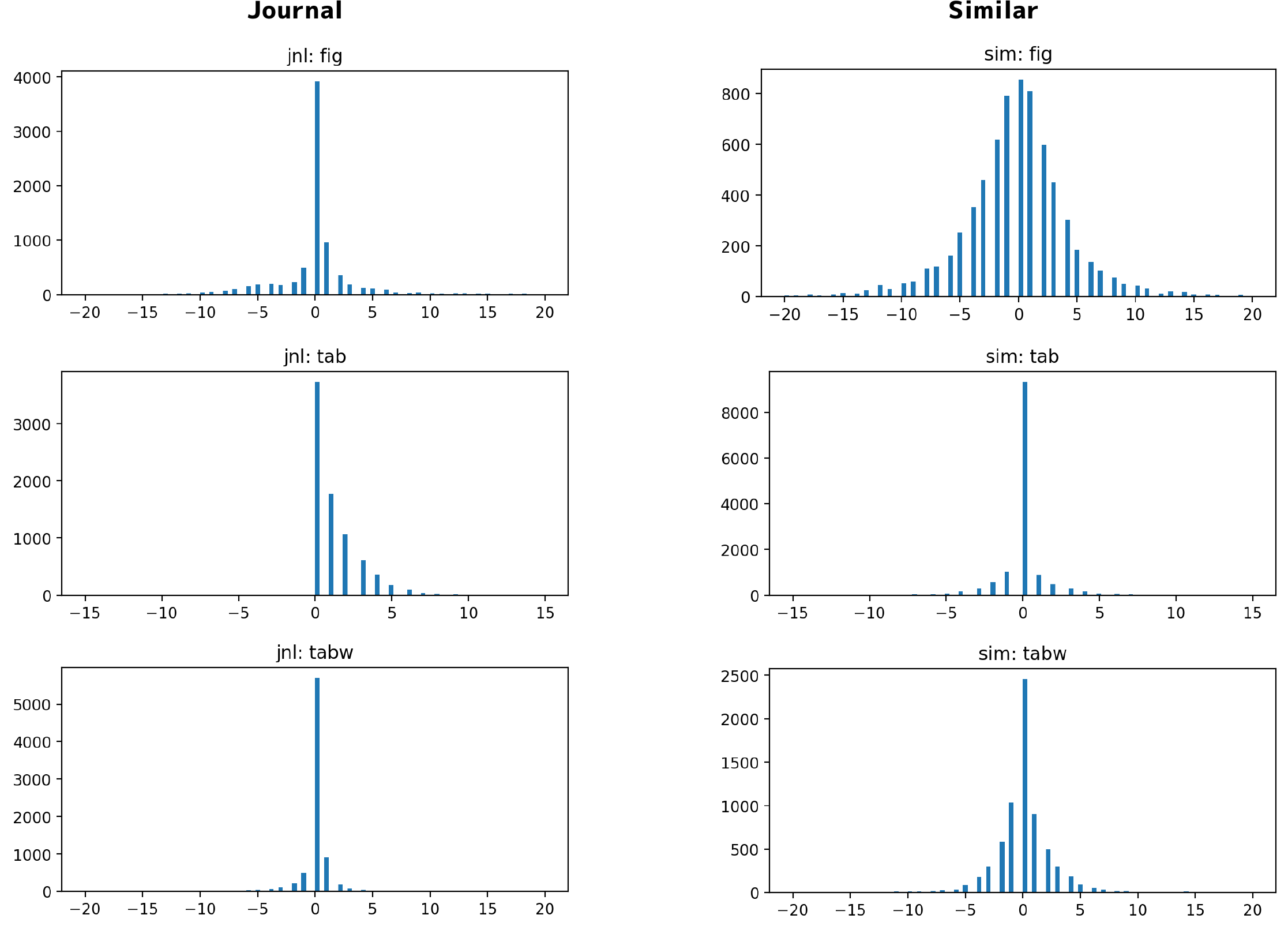}
\caption{Distribution of the difference in the number of charts}
\label{fig_fig_tab}
\end{figure*}

\begin{figure*}[htbp]
\centering
\includegraphics[width=140mm]{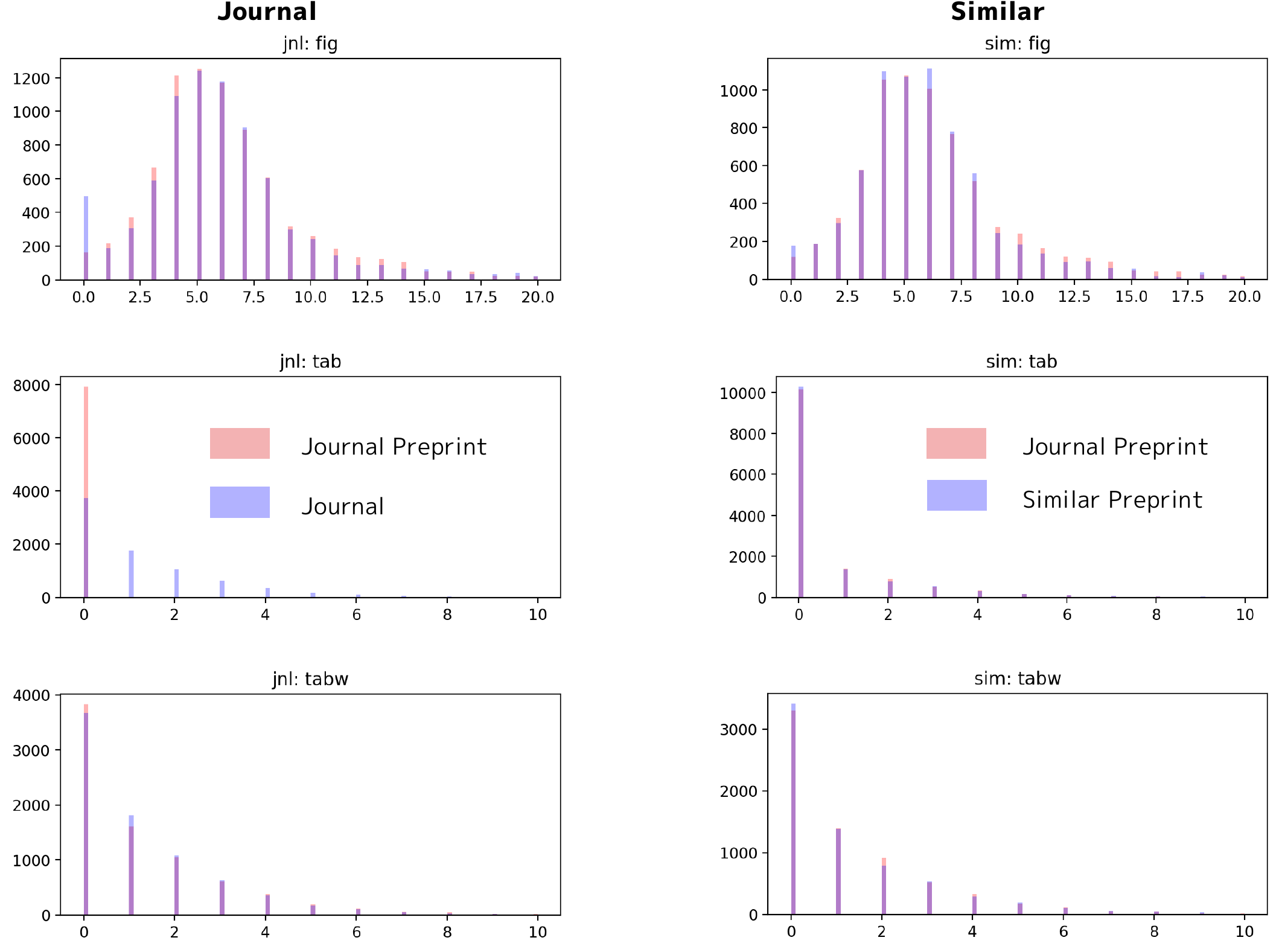}
\caption{Distribution of the number of charts}
\label{fig_fig_tab_num}
\end{figure*}

Figure \ref{fig_fig_tab},\ref{fig_fig_tab_num} shows the difference and the real distribution of the number of figures. In figures \ref{fig_fig_tab},\ref{fig_fig_tab_num}, tab refers to the table tag in JATS-XML, and tabw refers to the table-wrap tag. Since table tags are rarely used in preprints, and table-wrap tends to be used more frequently, it is necessary to pay attention to the handling of table and table-wrap.

\begin{figure*}[htbp]
\centering
\includegraphics[width=140mm]{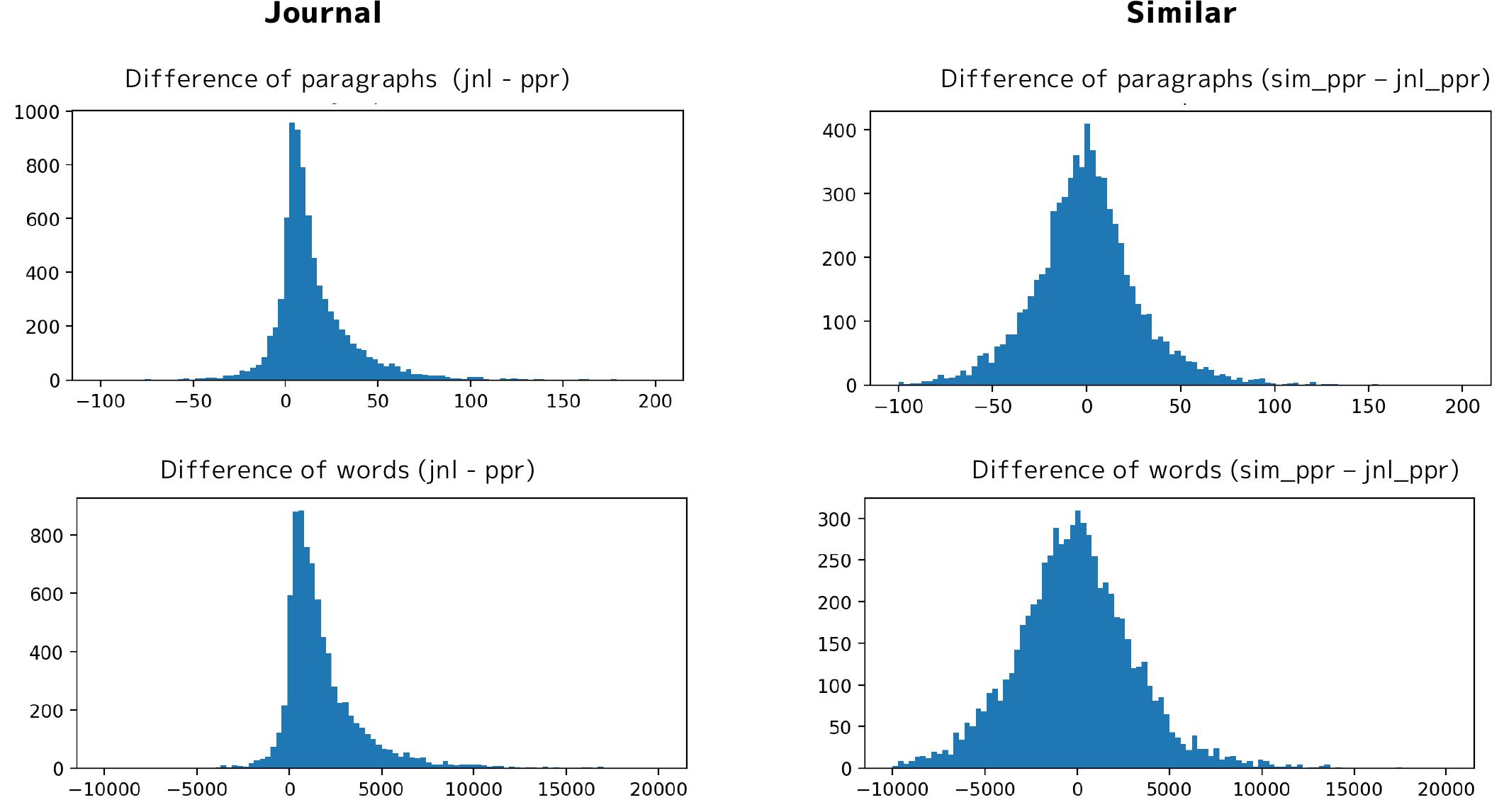}
\caption{Distribution of the difference in the number of paragraphs and words}
\label{fig_p_word}
\end{figure*}

\begin{figure*}[htbp]
\centering
\includegraphics[width=140mm]{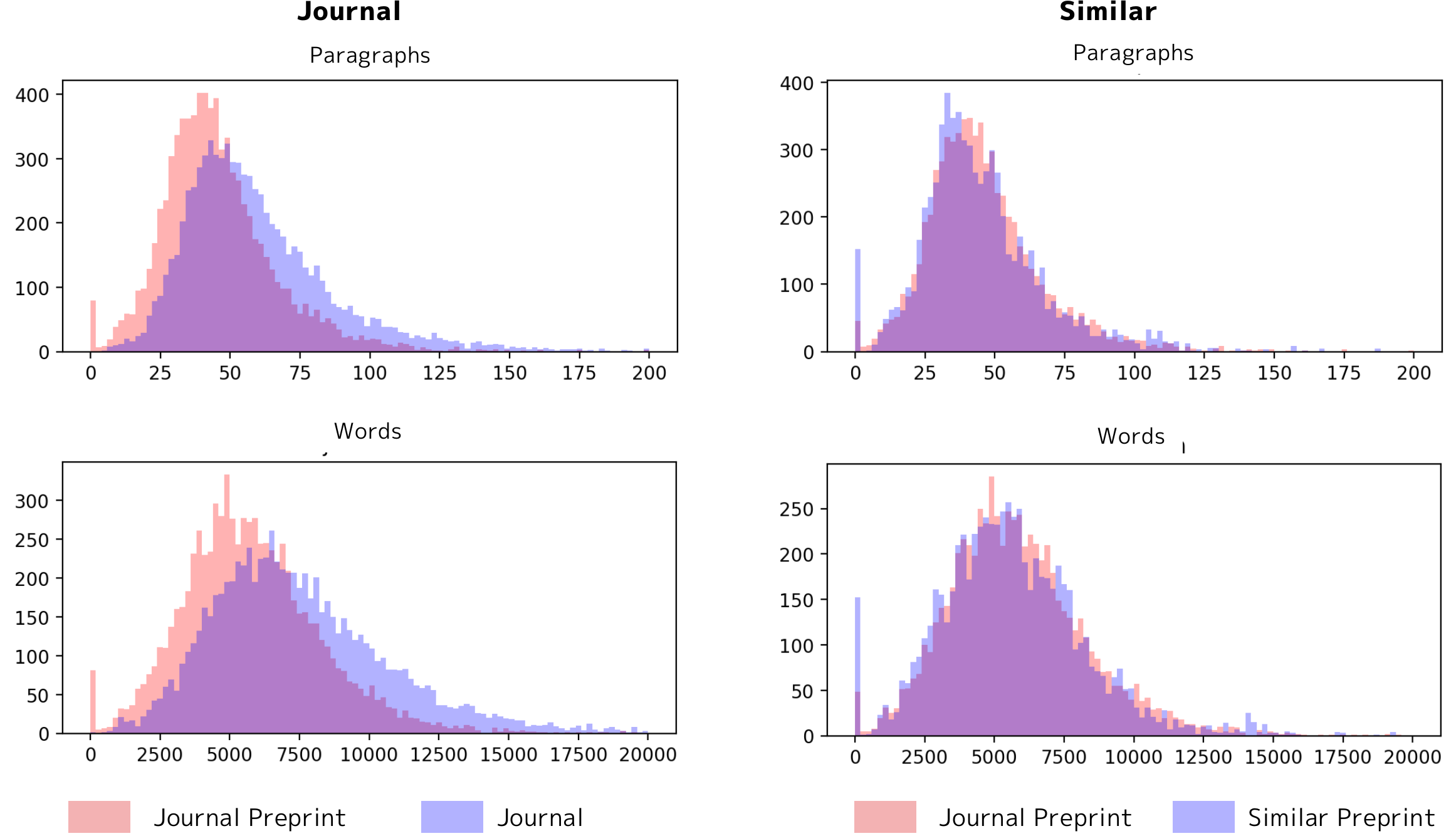}
\caption{Distribution of the number of paragraphs and words}
\label{fig_p_word_num}
\end{figure*}

Figure \ref{fig_p_word} and \ref{fig_p_word_num} show the difference in the number of paragraphs and words and the real number distribution. Figure \ref{fig_p_word_num} shows that the distribution of the number of paragraphs and the number of words shifted to the right when compared with that of the journal article, indicating that the number of paragraphs and the number of words increased in the journal article. Figure \ref{fig_p_word} also shows that the number of paragraphs and words rarely decreases.

\begin{figure*}[htbp]
\centering
\includegraphics[width=140mm]{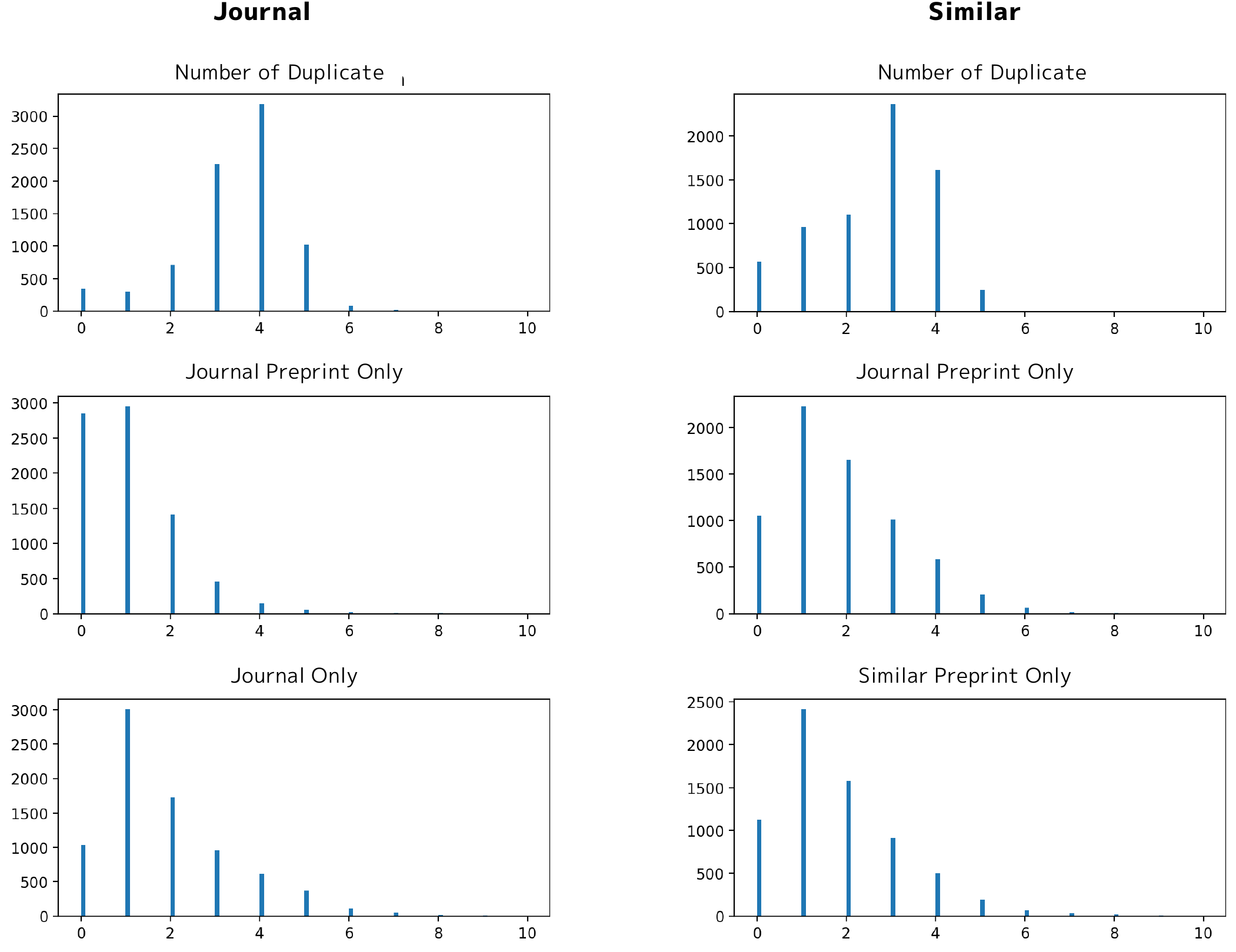}
\caption{Distribution of the sum and difference sets of the chapter title}
\label{fig_sectitle}
\end{figure*}

The distribution of the number of identical and different chapter titles is shown in Figure \ref{fig_sectitle}.

% ------------------------
\subsection{Investigation of differences in titles}

In addition to simply looking at the number of titles, we also looked at overlapping titles and different titles.

% +++++++++++
\subsubsection{Differences in chapter titles}

The differences in chapter titles were organized into a word cloud of common titles, titles that existed only in journal preprints, and titles that existed only in journal articles or similar preprints.

In this comparison, numbers such as ``1.'' in front of the title, periods and spaces at the end of the title were removed, and case was ignored.

\begin{figure*}[htbp]
\centering
\includegraphics[width=70mm]{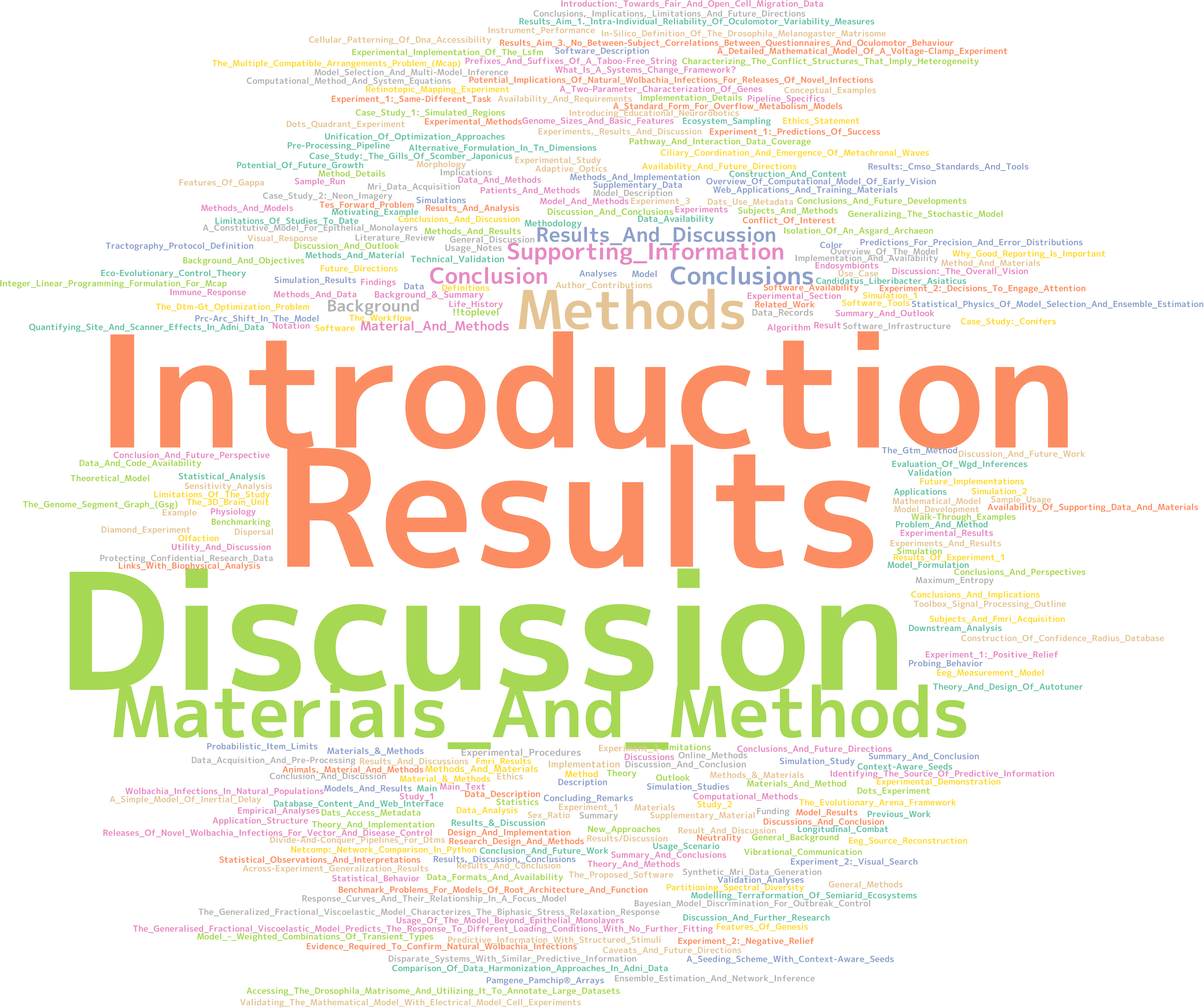}
\caption{Duplicates (Journals)}
\label{fig_jnl_sectitle_same}
\end{figure*}

\begin{figure*}[htbp]
\centering
\includegraphics[width=70mm]{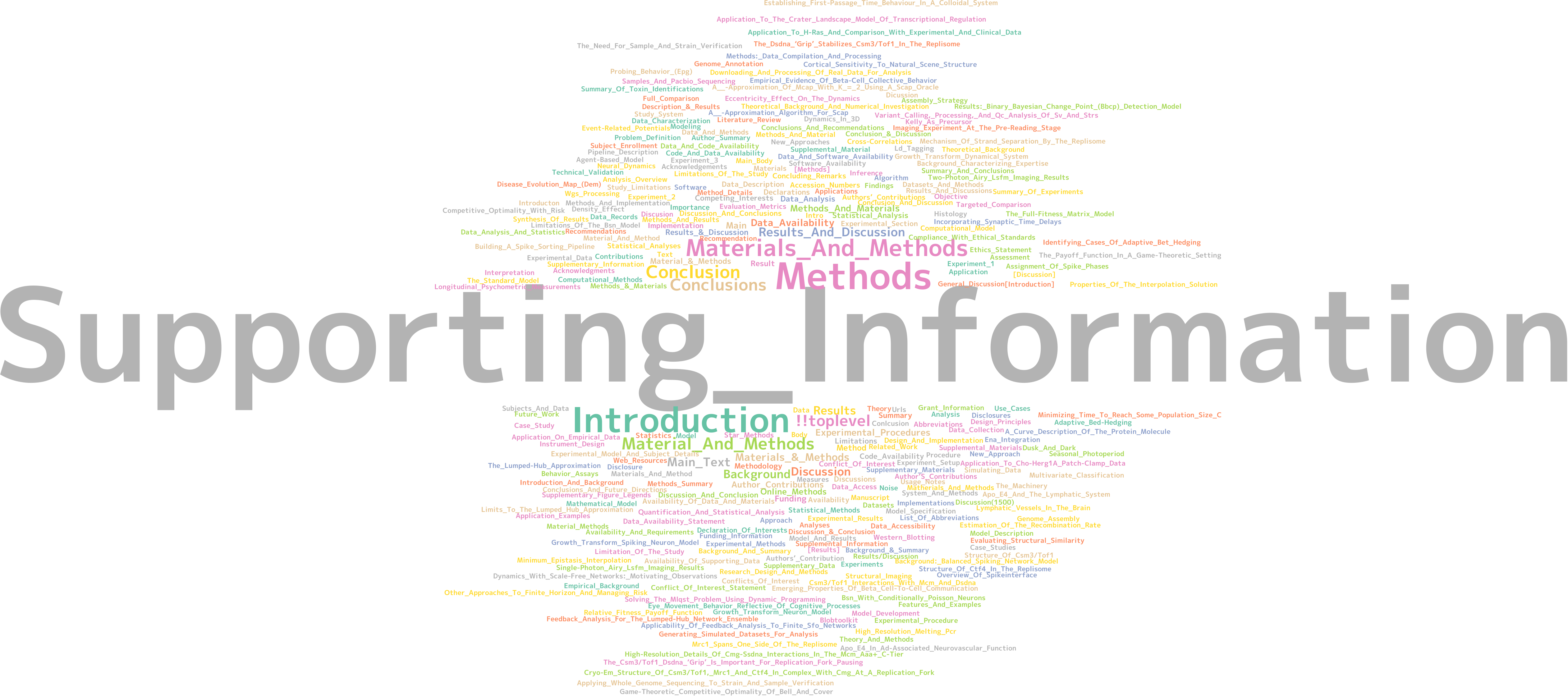}
\caption{Journal preprint only (Journal)}
\label{fig_jnl_sectitle_diff12}
\end{figure*}

\begin{figure*}[htbp]
\centering
\includegraphics[width=70mm]{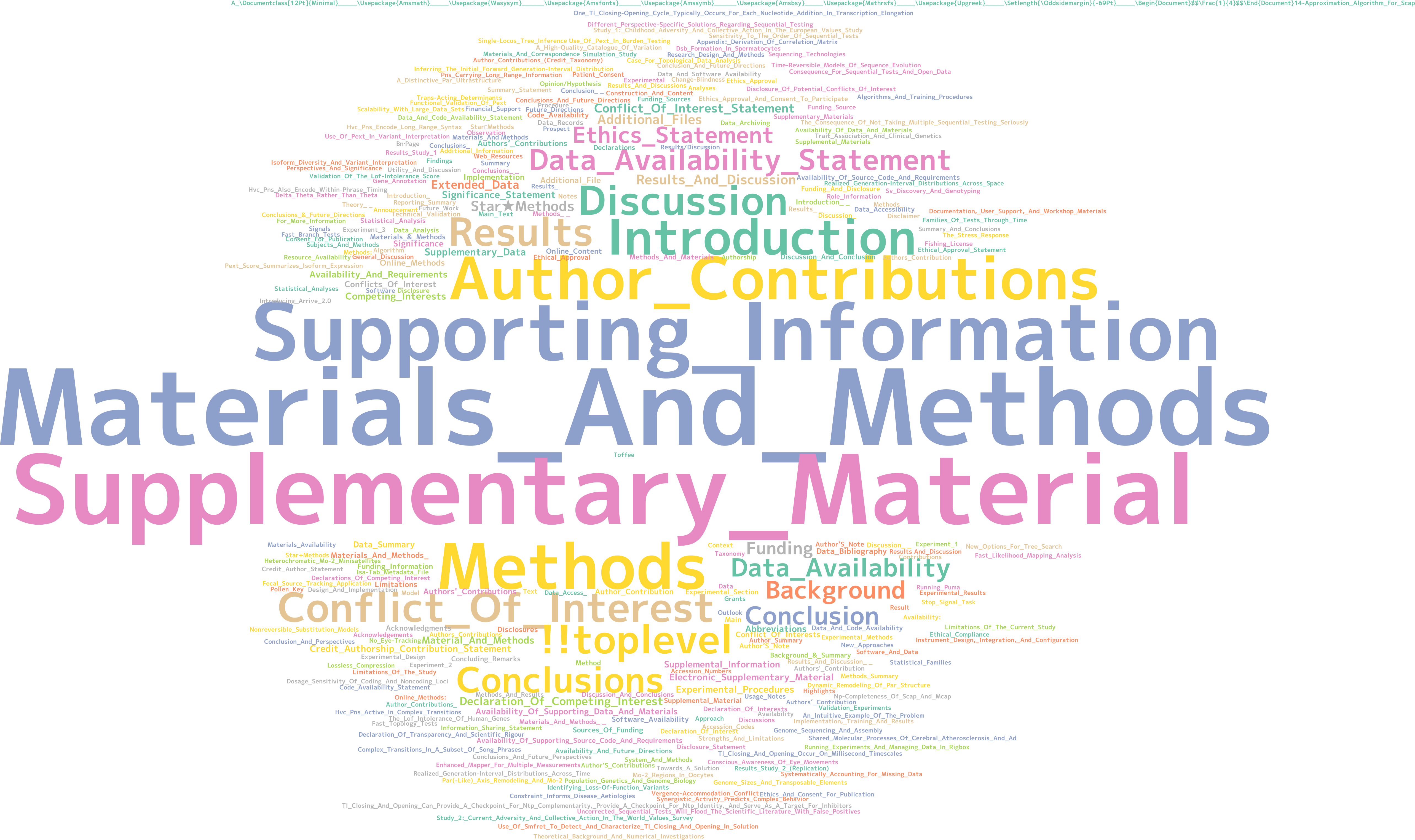}
\caption{Journal only (Journal)}
\label{fig_jnl_sectitle_diff21}
\end{figure*}

Figures \ref{fig_jnl_sectitle_same}, \ref{fig_jnl_sectitle_diff12}, \ref{fig_jnl_sectitle_diff21} show the results of the survey on journal preprints and journals.


Comparing the figures \ref{fig_jnl_sectitle_diff12},\ref{fig_jnl_sectitle_diff21}, we can see that the frequent words are similar, e.g. Material is deleted or added in some cases.

\begin{figure*}[htbp]
\centering
\includegraphics[width=70mm]{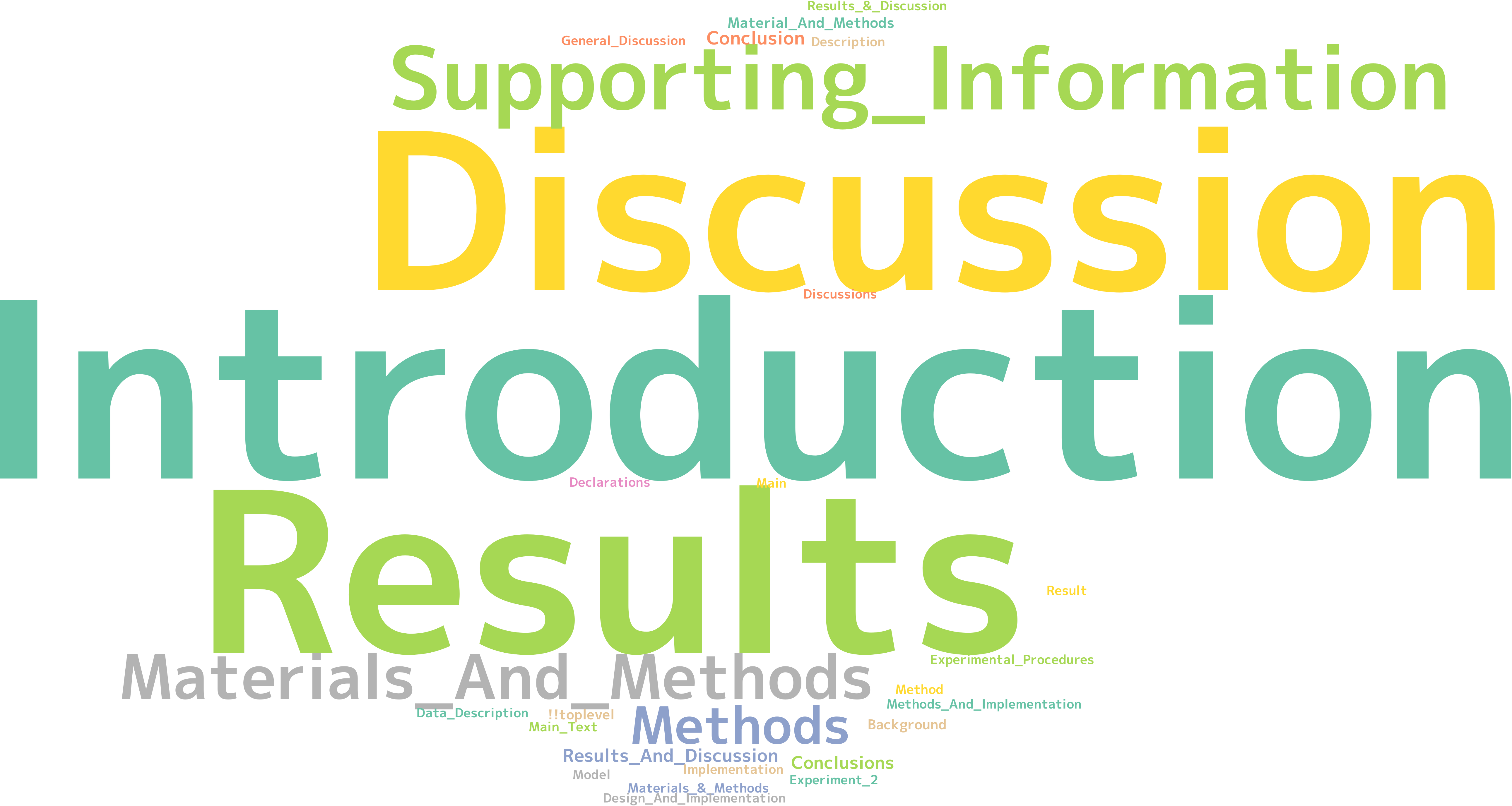}
\caption{Duplicates (similar)}
\label{fig_sim_sectitle_same}
\end{figure*}

\begin{figure*}[htbp]
\centering
\includegraphics[width=70mm]{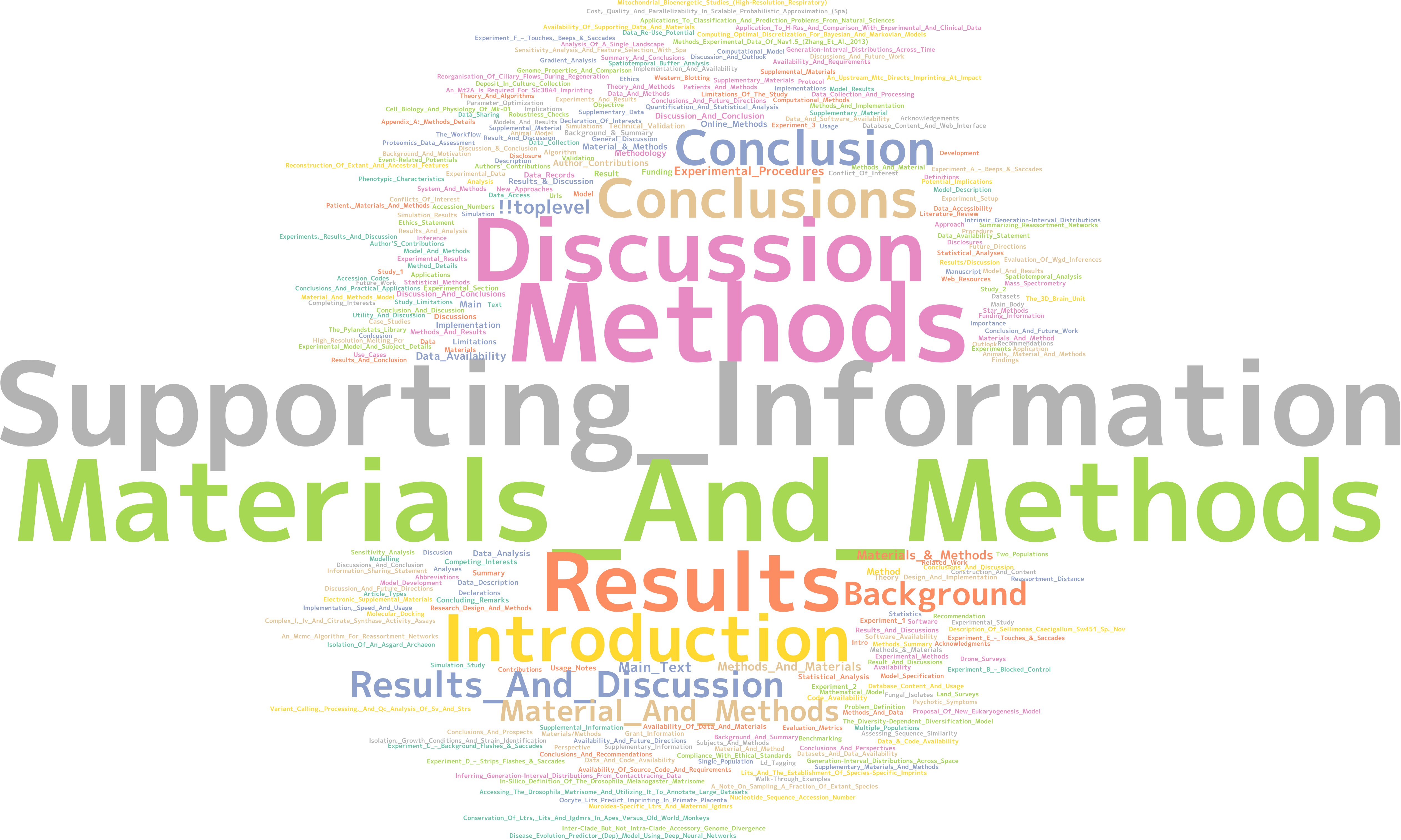}
\caption{Journal preprint only (similar)}
\label{fig_sim_sectitle_diff12}
\end{figure*}

\begin{figure*}[htbp]
\centering
\includegraphics[width=70mm]{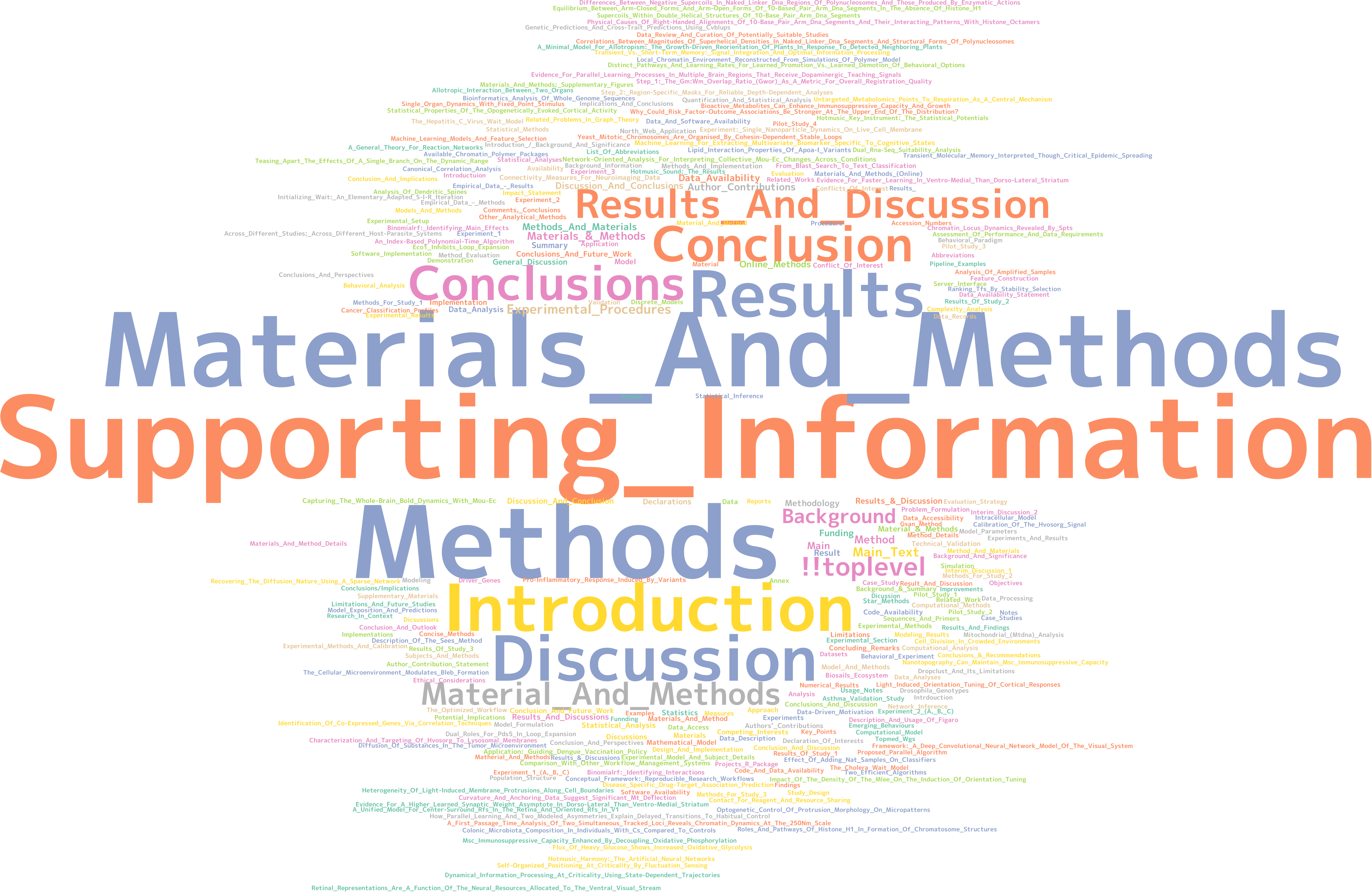}
\caption{Similar preprints only (similar)}
\label{fig_sim_sectitle_diff21}
\end{figure*}

Journal preprints - Figure \ref{fig_sim_sectitle_same},\ref{fig_sim_sectitle_diff12},\ref{fig_sim_sectitle_diff21} shows the results of the survey on similar preprints.


Comparing Figures \ref{fig_sim_sectitle_diff12} and \ref{fig_sim_sectitle_diff21}, we can see that the frequent words are similar, for example, Material is deleted or added in some cases. However, the figure \ref{fig_sim_sectitle_same} shows a bias toward certain words, indicating the existence of a chapter structure common to preprints.

% +++++++++++
\subsubsection{Title differences}

For journal preprints and journals, we also investigated the difference in the title of the article itself, not the chapter title.

Since we wanted to observe the changes in titles after peer review and proofreading, and since there are various variations in titles in general and duplication is not meaningful, we only investigated the degree to which the word sets of titles match (Jaccard) and the differences in specific words.

\begin{figure*}[htbp]
\centering
\includegraphics[width=140mm]{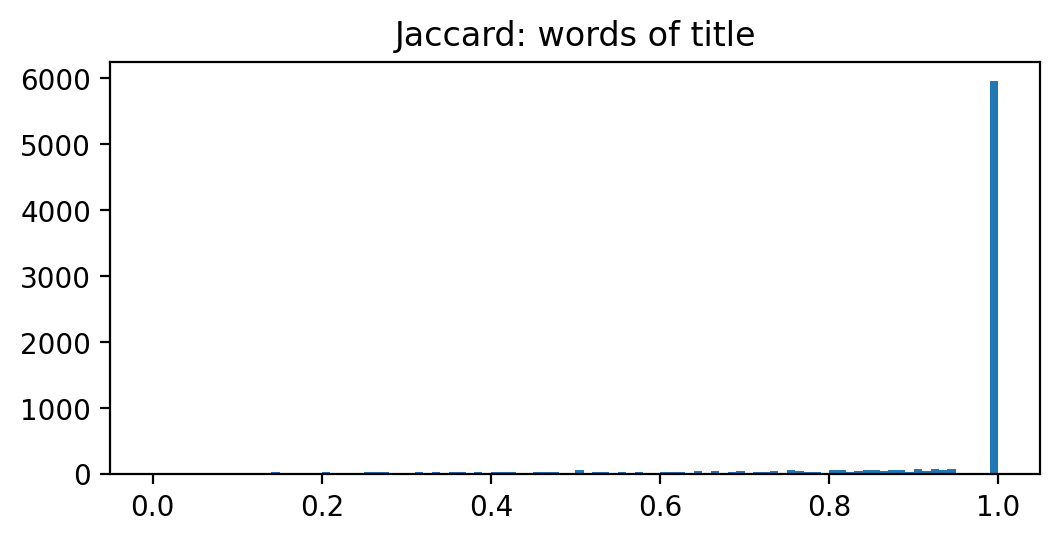}
\caption{Distribution of Jaccard coefficients}
\label{fig_title_dist}
\end{figure*}

\begin{figure*}[htbp]
\centering
\includegraphics[width=70mm]{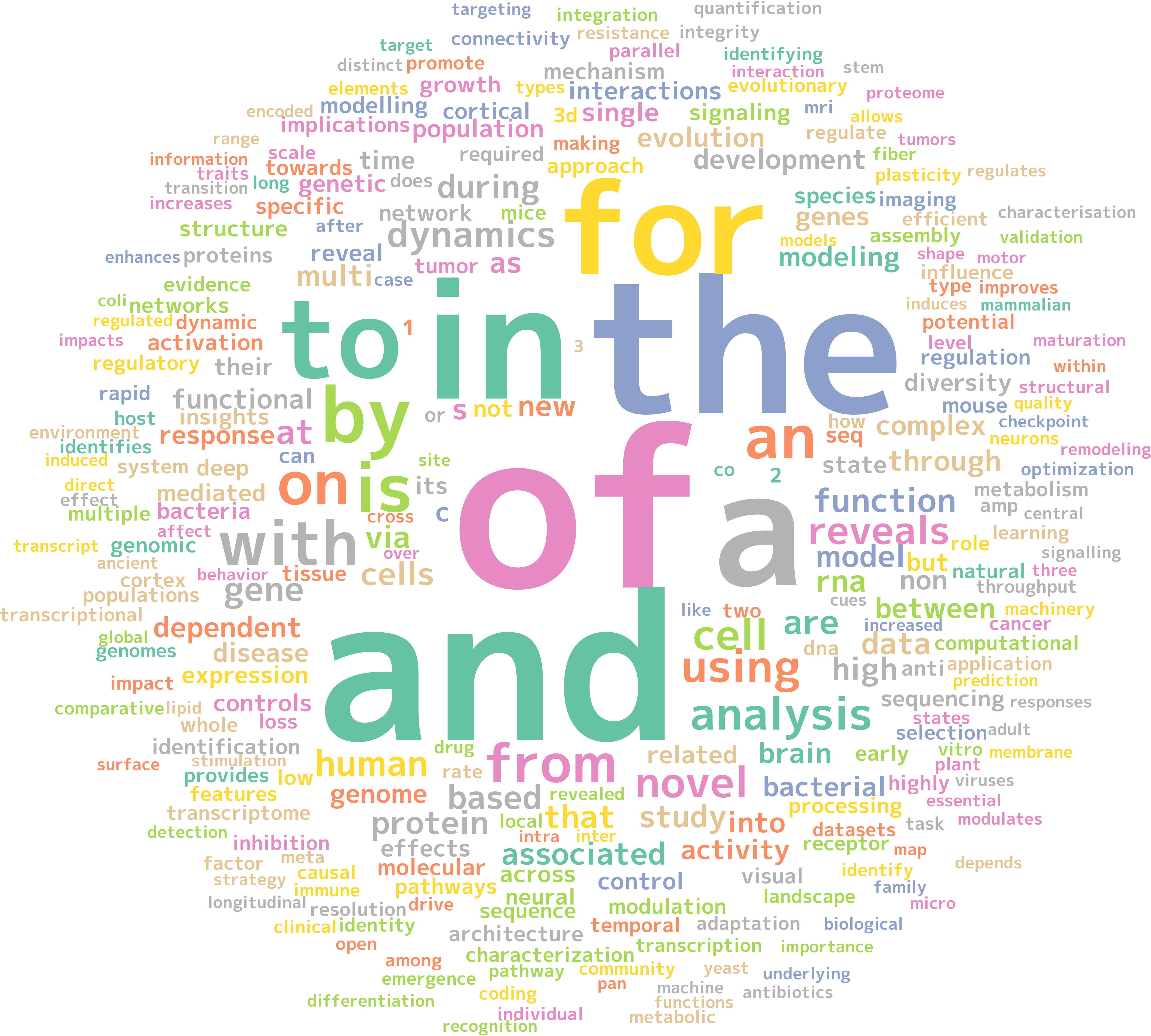}
\caption{Journal preprints only}
\label{fig_wc_d1}
\end{figure*}

\begin{figure*}[htbp]
\centering
\includegraphics[width=70mm]{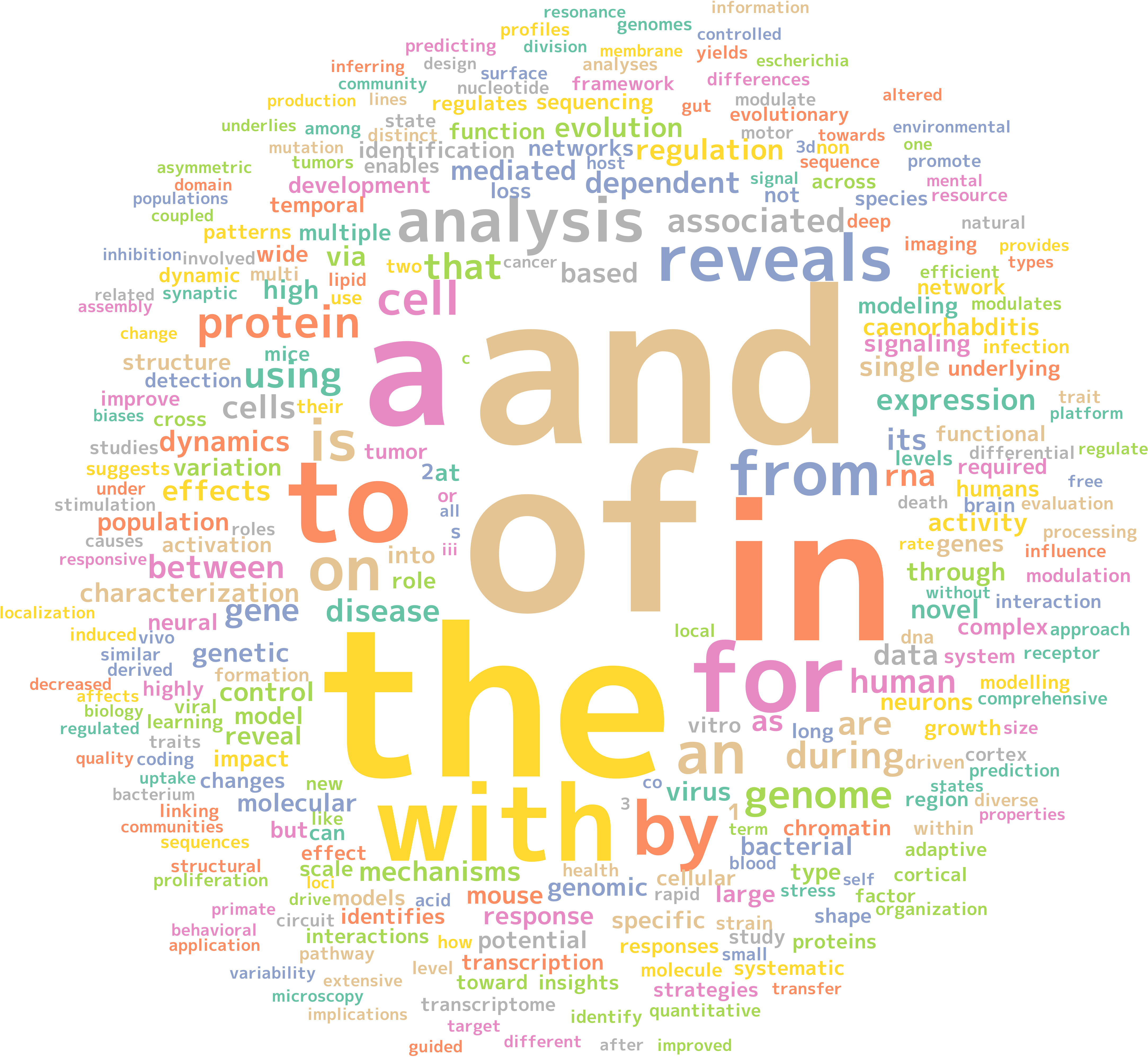}
\caption{Journals only}
\label{fig_wc_d2}
\end{figure*}

The results are shown in Figure \ref{fig_title_dist},\ref{fig_wc_d1},\ref{fig_wc_d2}.


Figure \ref{fig_title_dist} shows that there is no difference in the word set comparison in most cases (Jaccard coefficient is 1.0), and there is no significant difference in the concrete words between figures \ref{fig_wc_d1} and \ref{fig_wc_d2} as well as in the chapter titles. In other words, in some cases the definite article was added, in other cases it was removed, and so on, on a case-by-case basis.

% ------------------------
\subsection{Estimating the importance of external criteria}

So far, there was no significant difference between journal preprints and journal articles, although there were some differences in word count and number of references.

However, even if there is no difference in one indicator, there may be a difference in some conditions and combinations of several indicators. Therefore, we investigated whether we could distinguish journal articles, journal preprints, and other preprints using machine learning.

In this paper, we use the journal articles as they have been used in the past, but change the scope of the journal preprints and introduce a new category called ``other preprints''. These are measures to increase the sample size.

First of all, journal preprints used to include only those for which the full-text XML of the paired journal article could be obtained, but since the pair is not necessary here, we consider 12.925 preprints that are associated with journal articles and for which the full-text XML of the preprint can be successfully obtained, regardless of whether the full-text XML is available or not, Therefore, 12,925 items that were associated with journal articles with or without full-text XML and for which the full-text XML of the preprint was successfully obtained were considered as journal preprints. In addition, 14,673 cases that were not associated with a journal article and the full-text XML of the preprint was successfully obtained were classified as preprints.

For each of these types, we generated a classifier that classified them based on external criteria, and examined the importance of each criterion (feature). The following criteria were set in reference to the previous analysis.

\begin{table}[htbp]
    \centering
    \begin{tabular}{r|l}
\hline
Labels & Description \\
\hline
\hline
auth & Number of authors \\
ref & Number of references \\
word & Total number of words \\
fig & Number of figures \\
tab & Number of tables \\
intro & Percentage of total words in chapters with ``intro'' in the chapter title \\
metho & Percentage of chapter words that contain ``method'' in the chapter title \\
resul & Percentage of all chapter words that contain ``result'' in the chapter title \\
discu & Percentage of the total number of chapter words that contain ``discuss'' in the chapter title \\
concl & Percentage of all chapter words that contain ``conclusion'' in the chapter title \\
\hline
    \end{tabular}
    \caption{Features used for classification}
    \label{tab:charset}
\end{table}

The classification method used was RandomForest in the Python machine learning package scikit-learn, 
with 70\% of the data randomly selected as training data and 30\% as test data.

The results are shown below.

\begin{table}[htbp]
\centering
\begin{tabular}{rr|rrr}
\hline
                         &             & \multicolumn{3}{c}{\textsf{Exact}}                  \\
                         &             & Journal & Journal-PPr & PPr  \\
\hline
\hline
\multirow{3}{*}{\textsf{Predict}} 
                         & Journal     & 1,156    &   595        &   524 \\
                         & Journal-PPr & 1,003    & 1,922        & 1,863 \\
                         & PPr         &   834    & 2,695        & 3,487 \\
\hline
\multicolumn{5}{r}{* PPr: PrePrint}                                  
\end{tabular}
\caption{Classification results of test data by RandomForest}
\label{tab:predict}
\end{table}

Classification results of test data by RandomForest

\begin{figure*}[htbp]
\centering
\includegraphics[width=140mm]{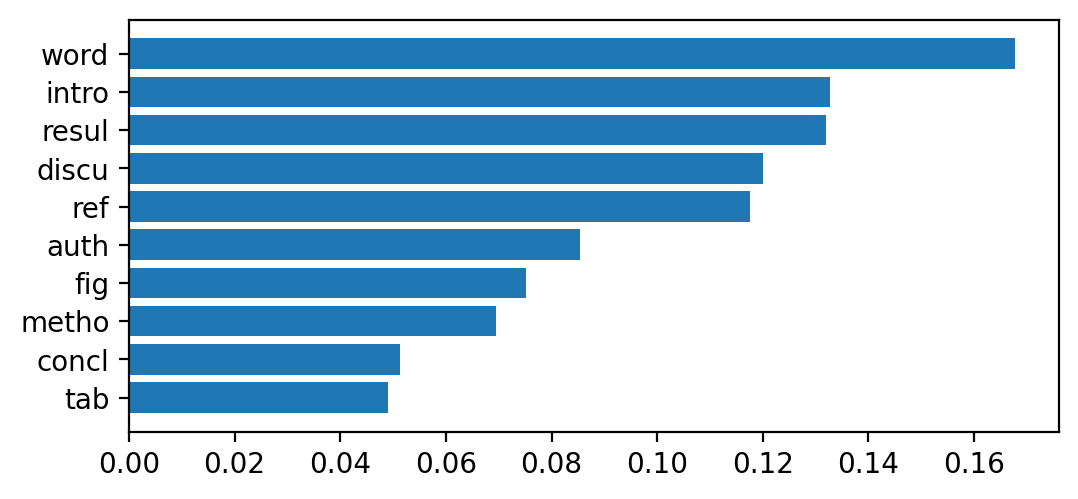}
\caption{Importance of features}
\label{fig_rf_importance}
\end{figure*}

The accuracy is 0.47, and the importance of the features is as shown in Figure \ref{fig_rf_importance}.

In the case where there is no prior knowledge, if the number of classifications is 3, the probability is simply 1/3, then the accuracy of 0.47 is higher than that, but it is not good. In addition, the number of words is the most important factor in discrimination, followed by the introduction and conclusion.

% ==================================================================
\section{Discussion}

From this analysis, we can see that there seems to be no difference between journal preprints, journal articles, and similar preprints in general. In other words, it would be difficult for a person who does not have expertise in the field to distinguish between a journal preprint, a journal article, and a similar preprint based on the chapter structure and the number of references.

In addition, the differences in content between journal preprints and journal articles are relatively small. However, even in the case of random matching, the peak value of the content similarity distribution of our method is high, about 0.93. In comparison with similar preprints, not a few of them have a similarity of 1.0. Thus, this criterion tends to estimate the similarity relatively high. In addition, the similarity calculation method truncates formulas and numbers, but in a paper, these small differences may be significant.

In addition, there are various points to be considered when reading the results, which will be discussed later, but to simplify the discussion, it is difficult to distinguish preprints and journal articles externally within the scope of this analysis. Therefore, it is necessary to conduct more in-depth text mining to answer the initial question, ``How are journal articles produced?'' and to examine the differences between preprints and journal articles by adding experts in the field.

% ------------------------
\subsection{Points to be noted}

This time, bioRxiv was selected as the preprint server. Therefore, the field is limited to biology only. In addition, although biological sciences are more active than other fields, the diffusion rate of bioRxiv in biological sciences is not high, considering the number of researchers in the field and the usage rate of arXiv in physical and information sciences. In addition to this, the number of journal articles we were able to survey was very small (about 7,000), and only open access articles were available. Therefore, it should be noted that the discussion is limited to Open Access articles of advanced users who use preprints in the biological sciences.

In addition, although some of the contents were analyzed, they were processed mechanically and the specific contents and contexts were not considered. In reality, even if the set of words is the same, the meaning may be completely different if the order changes, and even if the sentences are almost the same, small differences in numerical values may have critical meanings. These points should also be kept in mind.

The ``quality'' of a preprint and a journal article also requires careful discussion. It is not clear whether the quality and value of a paper is low because it has not been published in a journal. Some papers may have been accepted for publication simply because they have not been submitted to journals, while others may have been submitted but were not evaluated properly and remain as preprints because the reviewers' fields of expertise happen to be different. There may be some papers that are not evaluated at present, but will be of great value when combined with other results in a few decades. There are various possible reasons why a paper has not been published in a journal. In this paper, we have investigated the differences between preprints, journal articles, and preprints, but we have only investigated the differences and not how they affect the nature of the research. We did not clarify how critically the added references affected the content, how the changes in chapter titles and organization affected the content, or how much the value of the research changed as a result. It should be noted that the paper merely states that there is such and such a difference.

% ==================================================================
\section{Conclusion}

In this paper, we attempted to obtain knowledge about how research is conducted, especially how journal articles are produced, by comparing preprints with journal articles that are finally published.

First, due to the recent trend of open journals, we were able to secure a certain amount of full-text XML of preprints and journal articles, and verified the technical feasibility of comparing preprints and journal articles.

On the other hand, within the scope of this trial, in which we tried to clarify the difference between them based on external criteria such as the number of references and the number of words, and simple document similarity, we could not find a clear difference between preprints and journal articles, or between preprints that became journal articles and those that did not. Even with the machine learning method, the classification accuracy was not high at about 47\%.

The result that there is no significant difference between preprints and journal articles is a finding that has been shown in previous studies and has been replicated in larger and relatively recent situations.
In addition to these, the new findings of this paper are that the differences in many external criteria, such as the number of authors, are small, and the differences with preprints that are not journal articles are not large.

Thus, in order to verify how journal articles are produced, what points are brushed up from the preprint stage to become a journal article, and what are the critical differences between those that are accepted as journal articles and those that are not, it is necessary to include experts in the field and conduct more advanced research. In order to examine these issues, it was suggested that it is necessary to include experts in the field and conduct more advanced analysis using text mining and other methods.

As a derivative suggestion, preprints are externally equivalent to journal articles in terms of volume and other factors, and it is difficult to find out what is likely to be accepted for journal articles from preprints based on external criteria. In addition, there is a possibility that the necessity and significance of the peer review system can be reevaluated by pursuing the approach of this study.

%=============================================

\bibliographystyle{unsrtnat}

% --------------------------------------------------------------------
\clearpage
\appendix
\section{Comparison of the number of citations}
The number of citations is not used in this report because there are many issues to be solved in using it as an indicator, such as the evaluation differs depending on the timing of the measurement.

\begin{figure*}[htbp]
\centering
\includegraphics[width=120mm]{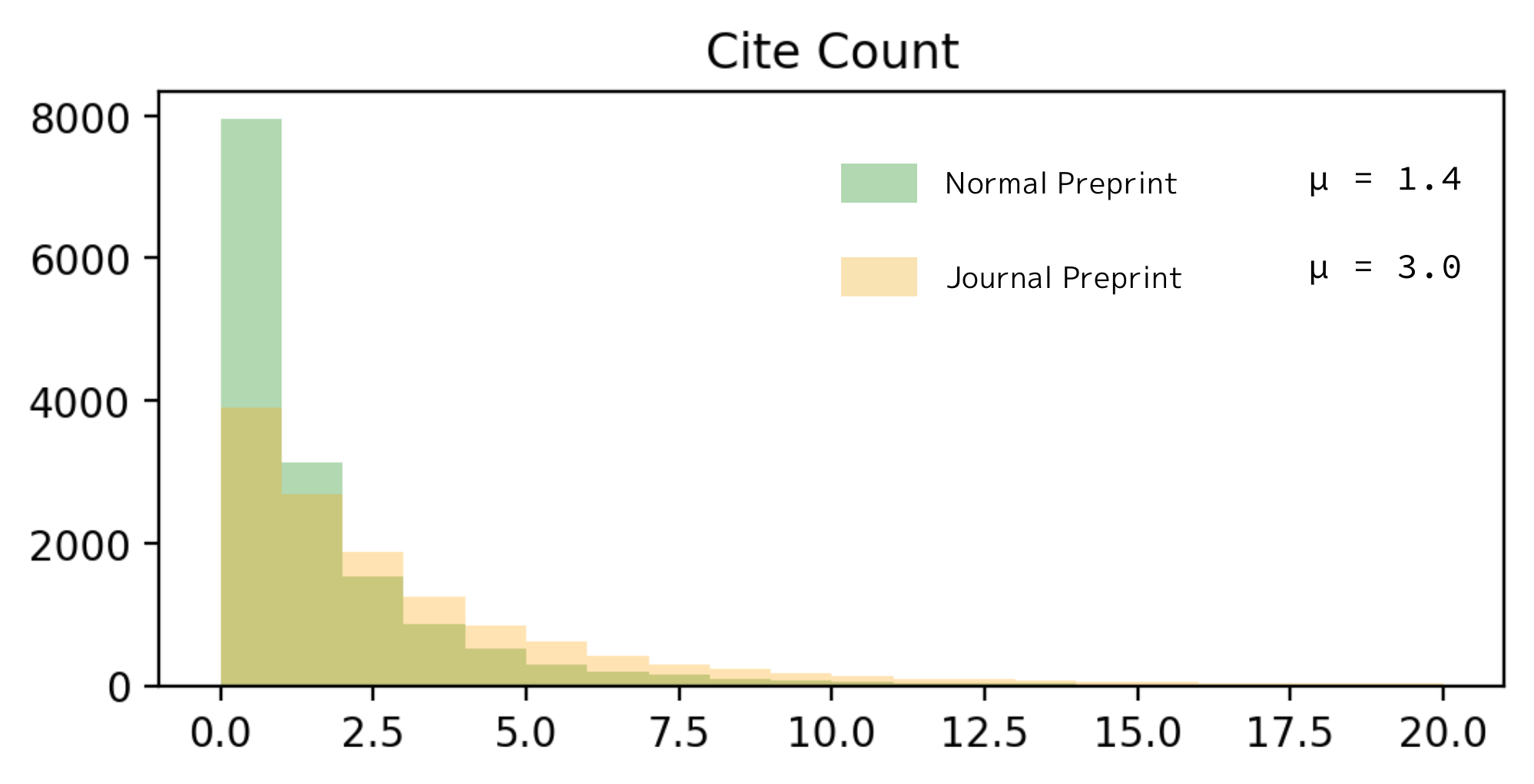}
\caption{Comparison of citations}
\label{fig_cite}
\end{figure*}

Figure \ref{fig_cite} shows a comparison of the number of citations between journal preprints and other preprints as of the end of April 2021.

The number of citations is higher for journal preprints.
It is unclear whether the journal preprints receive more attention and citations because they are published in a journal, or whether frequently cited articles are more likely to be published in a journal.

\section{Comparison of the number of versions}
The number of versions is not used in this report because there are many issues as same as citations, to be solved in using it as an indicator, such as the evaluation differs depending on the timing of the measurement.

Figure \ref{fig_cite} shows a comparison of the number of version counts between journal preprints and other preprints as of the end of April 2021.

\begin{figure*}[htbp]
\centering
\includegraphics[width=120mm]{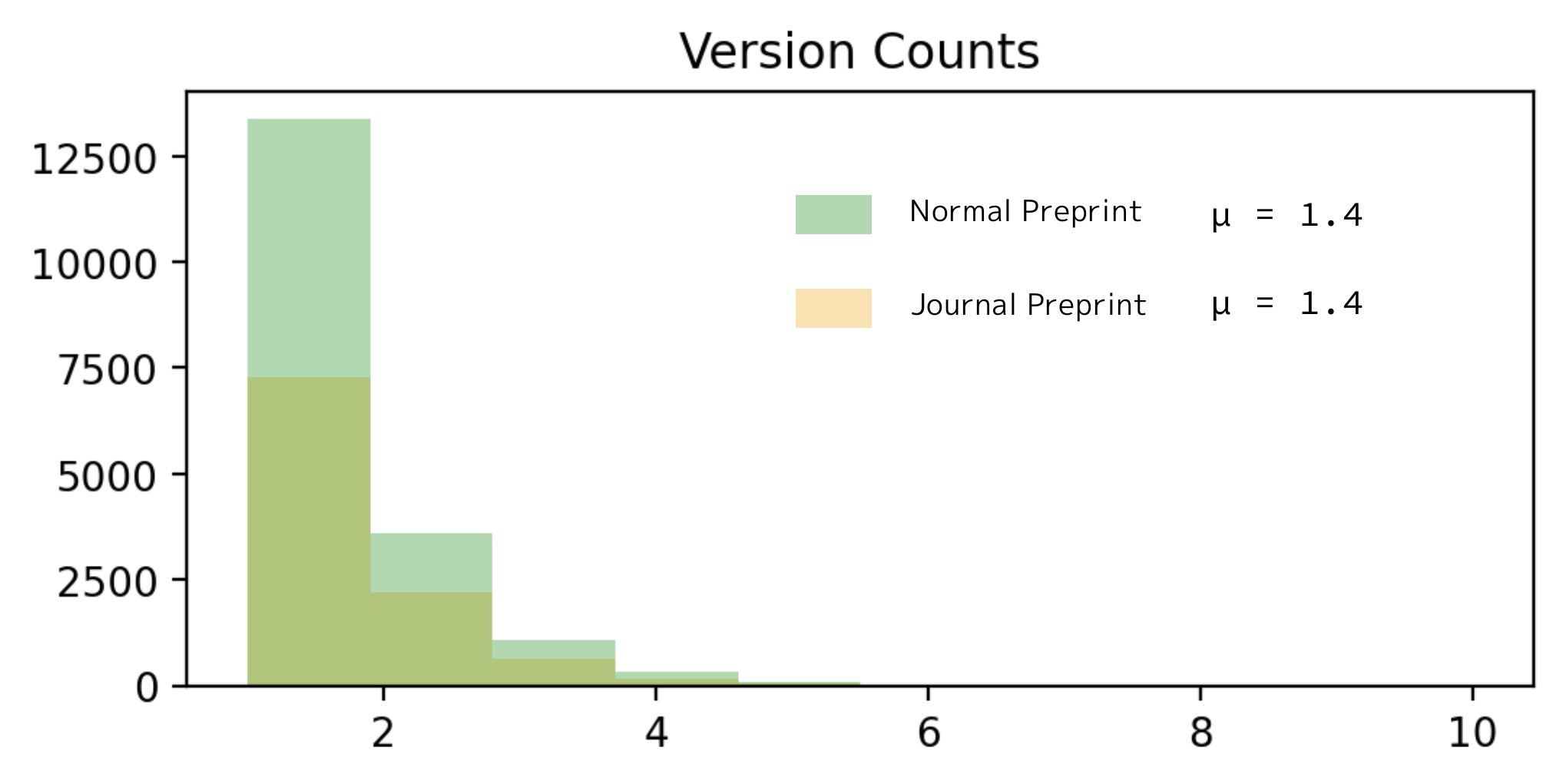}
\caption{Comparison of versions}
\label{fig_version}
\end{figure*}

There was no difference between journal preprints and other preprints, averaging 1.4 times, and the trends were consistent.

\end{document}